\RequirePackage{fixltx2e}
\documentclass[twocolumn,aps,pre,showpacs,amsmath,amsfonts,amssymb,floatfix]{revtex4}
\usepackage{morefloats}
\usepackage{amsmath}
\usepackage{graphicx}
\usepackage{dcolumn}
\usepackage{bm}
\usepackage{color}

\definecolor{orange}{rgb}{1,0.5,0}

\begin{document}

\title{Origin and implications of zero degeneracy in networks spectra}
\author{Alok Yadav}
\affiliation{Complex Systems Lab, Discipline of Physics,  Indian Institute of Technology Indore, Indore, Madhya Pradesh 452017, India}
\author{Sarika Jalan}\email{sarikajalan9@gmail.com}
\affiliation{Complex Systems Lab, Discipline of Physics,  Indian Institute of Technology Indore, Indore, Madhya Pradesh 452017, India}
\affiliation{Centre for Biosciences and Biomedical Engineering, Indian Institute of Technology Indore, Indore, Madhya Pradesh, 452017, India}
\date{\today}

\begin{abstract}
Spectra of real world networks exhibit properties which are different from the random networks. One such property is the existence of a very high degeneracy at zero eigenvalues. 
In this work, we provide possible reasons behind occurrence of the zero degeneracy in various networks spectra. 
Comparison of zero degeneracy in  protein-protein interaction networks of six 
different species and in their corresponding model networks sheds light in understanding the
evolution of complex biological systems.
\end{abstract}
\maketitle

Last two decades have witnessed a tremendous growth in the studies of complex systems under the graph theory framework \cite{Barrat_book}. 
This framework which describes a complex system in terms of its interacting units, has not only enabled us to understand the properties of large complex systems, but has 
also shed light on the dynamics of evolution of these structural properties \cite{barabasi2002,newman2003,strogatz2001,boccaletti2006}. 
Though spectral graph theory is a well established domain \cite{Van_book,Chung_book,Rowlinson_book}, 
most of the studies in complex systems research pertain to 
understanding the properties and behaviour of a system by analysis of various structural measures, whereas spectral analyses studies of 
graphs generated for real world systems are comparatively limited \cite{Farkas_PRE_2001,Newman_PRL_2012,SJ_plos,phya2014}. 
The spectral investigations indicate that the patterns of density distributions are distinguishing features of different classes of model networks \cite{Goh}. Further, extremal eigenvalues have been shown to contain useful information about the structure of the graphs \cite{Newman_SIAM_2003}. Although the bulk of real world networks bear reasonable similarities with the model networks \cite{Chung_book}, some properties differ significantly. One such property is the degeneracy at the zero eigenvalue. Almost all biological and technological networks exhibit high degeneracy at the zero eigenvalue \cite{Dorogovtsev_PRE_2003,Aguiar,phya2014}, 
gene duplication being one of the suggested reasons behind the occurrence of high degeneracy at zero eigenvalue in biological systems \cite{Kamp_PRE_2005}.
Many biological systems are known to follow gene duplication as the basic mechanism behind their growth \cite{Teichmann}, and 
from a very simple matrix algebra calculation we know that a node duplication leads to lowering of the rank of the corresponding matrix and hence contributing to one additional zero eigenvalue in the spectra. Although node duplication provides a clue to the origin of zero degeneracy \cite{Jost}, it fails to provide a quantitative measure of actual degeneracy 
observed in real world networks \cite{Dorogovtsev_PRE_2003}, indicating the contribution from other factors. Scale-free behaviour or sparseness of real world networks have been 
argued out to be other reasons responsible for degeneracy at the zero eigenvalues \cite{Dorogovtsev_PRE_2003,Aguiar,phya2014}. In this work, we  explore origin of zero eigenvalues in various model networks. We substantiate the results by considering various real world networks.

A network can be represented in terms of adjacency matrix which is defined as,\\
$A_{\mathrm {ij}} = \begin{cases} 1~~\mbox{if } i \sim j \\
0 ~~ \mbox{otherwise} \end{cases}$.

The eigenvalues of the adjacency matrix are denoted by $\lambda_i,i=1,2, \hdots, N$ such that $\lambda_1<\lambda_2< \lambda_3< \hdots < \lambda_N$.
A theorem \cite{linearBook} relating the degeneracy at zero eigenvalues with the properties of the matrix states that for an adjacency matrix of size $N$ and rank $r$ there will be exactly $N-r$ zero eigenvalues . Therefore, if we know the rank of an 
adjacency matrix, we can find out the degenerate zero eigenvalues. Factors responsible for lowering of the rank of an adjacency matrix are enlisted in the following:
 
(a) When two rows (columns) have exactly same entries, it is termed as complete row (column) duplication:
\begin{equation}
R_1=R_2
\label{complete}
\end{equation}
Subtracting one such row from the other yields one of the rows to attain all zero values, thus reducing the rank of the matrix by one.

(b) When two or more rows (columns) added together have exactly same entries as some other row or column, we call it partial row (column) duplication. For example; 
\begin{equation} 
R_1=R_2+R_3, \, \, \, {\text or} R_1+R_2=R_3+R_4+R_5
\label{partial}
\end{equation}

(c)An isolated node in the network leads to all zero entries in the corresponding row and column, thus lowering the rank of the matrix by one.
\begin{table*}[t]
\begin{center}
\caption{In order to demonstrate one to one relation between number of duplicates and zero eigenvalues, each time one new add is added in a network of size $N$ in such a manner it satisfies either complete duplication criteria (ii) or the partial duplication criteria (iii). $D_c$ represents number of duplicate nodes and $\lambda_0$ indicates number of zero eigenvalues. Seed network of size $N=100$ and average degree $\langle k \rangle = 10$.}
\begin{tabular}{|c|c|c|c|c|c|c|c|c|c|c|c|c|c|c|c|c|}    \hline
\hline 
$N$ & 100 &101 &102 &102 &103 & 104 &104 & 105 &106 &106 & 107 & 108 &  108 & 109 & 110 & 110 \\ 
\hline 
$D_{c}$ & 0 & 1 & 2& 0&3&4&0&5&6&0&7&8&0&9&10 &0\\ 
\hline 
$D_{p}$ &0&0&0&1&0&0&2&0&0&3&0&0&4&0&0&5\\ 
\hline 
$\lambda_0$ &0&1&2&1&3&4&2&5&6&3&7&8&4&9&10&5\\ 
\hline 
Condition&-&a& a& b& a& a& b& a& a& b& a& a& b& a& a& b\\
\hline
\end{tabular}
\label{table1}
\end{center}
\end{table*} 

Conditions (a) and (b) lead to linear dependence of row (column), reducing the rank of the matrix.
Note that we consider a connected network in order to rule out the trivial possibility (c) of occurrence of zero eigenvalues. Further there are $N(N-1)/2$ possible ways in which condition (a) of complete duplication can be realized, while for the partial duplication (b) among `$x$' number of nodes with `$y$' number of nodes, there can be $\frac{N!}{2.(N-x-y)!}$ possibilities. Hence, for a given network, checking the existence of condition (b) becomes computationally exhaustive as with increase in network size the number of possibilities becomes very large. 
\begin{figure}[b]
\includegraphics[width=\columnwidth]{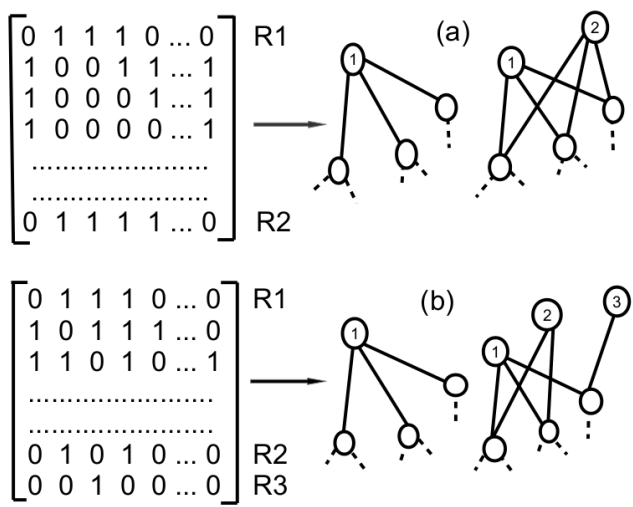} 
\caption{(Color online) Schematic diagram
representing (a) complete node duplication (Eq.~\ref{complete}) and (b) partial node duplication (Eq.~\ref{partial}) in networks.}
\label{duplication}
\end{figure}

\begin{figure}[t]
\includegraphics[width=\columnwidth]{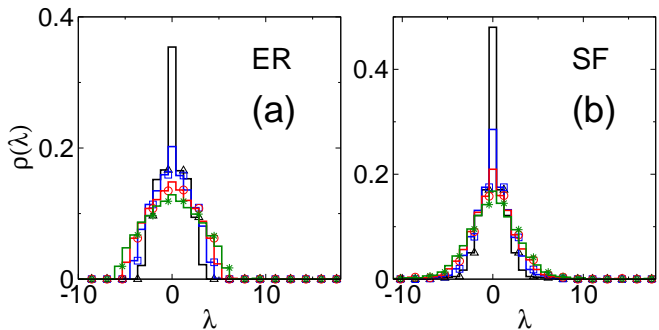}
\caption{(Color online) The density distribution of Erd\"os-Ren\'yi (ER) random networks and scale-free (SF) networks for different average degrees and $N=1000$. {\color{black} $\vartriangle$}, {\color{blue} $\square$}, {\color{red} $\circ$} and {\color{green} $\ast$} represent the data points of density distribution for $\langle k \rangle$=2, 4, 6 and 8, respectively. All values are averaged over 10 realizations of the networks.}
\label{lambdaER_SF_k}
\end{figure}

 In order to demonstrate the effect of duplication on zero degeneracy, we construct an Erd\"os-Ren\'yi (ER) random network for size $N$ and connection probability $p$ using ER model \cite{barabasi2002} such that it has no duplicates and no zero eigenvalues (row 1 of Table~\ref{table1}). Next we add a node to the existing network in a way that it satisfies the complete node duplication criteria, i.e. condition (a). This leads to exactly one zero eigenvalue corresponding to one duplicate node. Addition of one more node mimicking the previous node leads to two zero eigenvalues (row 2 and 3 of Table~\ref{table1}). This demonstrates how complete node duplication leads to zero eigenvalues (Fig.~\ref{duplication} (a)). Further, we consider another situation where we devise our algorithm such that two new nodes are added to the existing random network in a way that in coalition they mimic the neighbors of an existing node (condition (b)), i.e. they duplicate an existing node (row 4 of Table~\ref{table1}) as demonstrated in Fig.~\ref{duplication} (b). Impact of duplications of conditions (a) and (b) on the zero eigenvalues are presented in the subsequent rows of Table~\ref{table1}. Thus, we observe that with entry of every new node in the network satisfying condition (a) or (b) of complete or partial duplication, there is an addition of exactly one zero eigenvalue in the spectra. The number of duplicates (complete or partial) equals the number of zero eigenvalues. The density distribution at very low average degree yields a peak at zero eigenvalue. With an increase in $\langle k \rangle$, the peak of the density distribution flattens (Fig.~\ref{lambdaER_SF_k} (a)). 

In order to demonstrate the impact of network architecture on the duplication phenomenon, we present results for ensemble average of the scale-free (SF) networks as they are known to have high degeneracy at zero eigenvalue. We generate the SF network using the preferential attachment mechanism \cite{barabasi1999}, where each new node gets attached to the existing nodes with the probability proportional to their respective degrees. This phenomenon gives rise to power law degree distribution. Here at each time step, a new node enters which is most likely to connect with the highest degree nodes owing to the preferential attachment algorithm. The next entry also has a tendency to attach with the highest degree nodes. From the power law degree distribution of SF networks it is evident that there are very few high degree nodes which are known as the hubs of the network  and a large number of low degree nodes. At low values of $\langle k \rangle$, there is a high degeneracy at zero eigenvalue indicating high duplication. This is because at low  average degree, most of the low degree nodes attain very few connections. By virtue of preferential attachment property, these low degree nodes have the highest probability to connect with the hubs of the network, which increases the likelihood of any two nodes to have the same neighbors, leading to a pair of duplicate nodes. Although even with increase in average degree, the density distribution remains triangular, there is flattening of the peak (Fig.~\ref{lambdaER_SF_k} (b)). This might be because 
the low degree nodes also tend to acquire connections with nodes other than the hubs. All these findings indicate that low average degree favors duplication. The explanation behind this can be given in terms of the possible number of ways of duplication which is the ratio of the possible number of combinations of duplication possessed by two $k$-degree nodes to the possible number of combinations of random connections of those nodes. This is given as $\frac{k\,!}{N^{k}}$, where $N$ is the total number of nodes. As $k$ increases, the possible number of ways of duplication drastically decreases, thus explaining why low degree supports duplication.

\begin{figure}[b]
\includegraphics[width=\columnwidth]{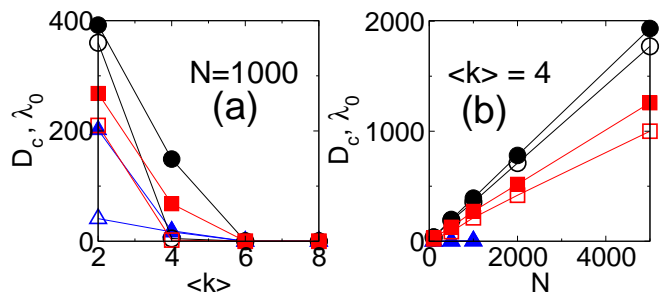}
\caption{(Color online) Effect of the change in the network parameters, namely, size ($N$)and the average degree ($\langle k \rangle$) on the number of the duplicates and zero eigenvalues in different model networks. All values are averaged over 10 random realizations of the networks. Note that for ER random networks, connected component could not be obtained at $\langle k \rangle$=4 for $N$ above 1000. Here the {\color{blue} $\vartriangle$}, {\color{black} $\circ$} and {\color{red} $\square $} represent the $D_c$ of the ER, SF and configuration model networks, respectively. The solid {\color{blue} $\vartriangle$}, {\color{black} $\circ$} and {\color{black} $\square$} represent the $D_c$ of the ER, SF and configuration model networks, respectively.}
\label{0_vs_model}
\end{figure}

Since networks with power law degree distribution, i.e. SF networks lead to high degeneracy in the spectra, we further explore other model networks with power law degree sequence. We construct the configuration model network by taking the degree sequence of the connected SF network as input. Each node of the corresponding configuration model is allotted stubs equal to their degree, then these stubs are paired with uniform probability \cite{Molloy,Newman2001,DelGenio2010}. This generates a configuration model for a given degree sequence. Only connected networks are carried further, rest all are discarded. In spite of having randomly assigned connections, they display a much higher zero degeneracy as compared to the ER random networks (Fig.~\ref{0_vs_model} (a)). The configuration model networks are not generated using the preferential attachment property as of the SF networks. So this possibility is ruled out as a reason behind the higher degeneracy of configuration model networks as compared to ER random networks. The particular (the power law) degree sequence emerges as a probable reason behind high degeneracy at zero eigenvalue in the configuration model. Due to the power law behaviour, there 
exists a large proportion of nodes with a low degree which for $\langle k \rangle$=2 are peripheral nodes. Only a few nodes having high degree act as the hubs. The large number of low degree nodes get randomly attached to high degree nodes i.e. the peripheral nodes attach with the hubs only, leading to the complete duplication of these nodes. While in case of ER networks, duplication is less likely as all the nodes have their degrees fluctuating around the average degree. The configuration model networks exhibit lower peak at zero eigenvalue as compared to SF networks (Fig.~\ref{0_vs_model}) as it is a randomized version of the SF network. This indicates that apart from the preferential attachment phenomenon, the particular degree sequence is also responsible for high degeneracy at zero eigenvalue.

\begin{table*}[ht]
\begin{center}
\caption{Properties of the six PPI networks and their comparison with ER random networks, SF networks generated using BA algorithm and corresponding configuration model networks (of the same degree sequence as of the PPI networks). $D_c$, $D^{ER}_{c}$, $D_{c}^{BA}$ and $D^{conf}_{c}$ denote the number of complete duplicates in the PPI networks, ER random networks, SF networks and configuration model networks, respectively. $\lambda_0$, $\lambda^{ER}_0$, $\lambda_0^{BA}$ and $\lambda^{conf}_{0}$ represent the number of degenerate zero eigenvalues in the PPI networks, ER random networks, SF networks and configuration model networks, respectively. The ER and SF networks are generated for average over 10 different realizations of the networks by keeping $N$ and average degree same.}

\begin{tabular}{|c|c|c|c|c|c|c|c|c|c|c|c|c|}    \hline

\hline 

Species & $N$ &$\langle k\rangle$& $\lambda_0$ & $D_{c}$ &$\lambda^{ER}_{0}$ &$D^{ER}_{c}$&$\lambda_0^{BA}$&$D_{c}^{BA}$&
$\lambda^{conf}_{0}$&$D^{conf}_{c}$\\ 

\hline 

{\it H. pylori} & 709 & 3.935&317&146 & 0&0&155&17&163&79\\ 

\hline  

{\it H. sapiens} & 2138 &2.872& 976&662 & 0&0&469&33&512&309\\ 

\hline 

{\it E. coli} & 2209 &9.895& 487 &323 & 0&0&0&0&569&200\\ 

\hline 

{\it C. elegans}& 2386 & 3.206&1354&940 & 0&0&528&29& 818&569\\ 

\hline 

{\it S. cerevisiae} & 5019 &8.803& 864&491 & 0&0&0&0&950&314\\ 

\hline 

{\it D. melanogaster} & 7321 &6.159 &2311&1046 & 0&0&110&0&1621&687\\ 

\hline  

\end{tabular}

\label{table2}

\end{center}

\end{table*}

So far we have discussed the impact of network architecture on the duplication and the zero degeneracy. In the following, we evaluate the impact of average degree and size on the same. For a fixed network size, when the average degree of the ER random network increases, there is an increase in the number of connections. The probability that any node has exactly the same set of neighbors as any other node is given by $\frac{\langle k \rangle^{2k}}{e^{2\langle k \rangle} k\,! N^k}$. With an increase in 
$\langle k \rangle$, this probability
 diminishes exponentially. This implicates reduction in the node duplication. Fig.~\ref{0_vs_model} (a) exhibits that at low average degree, the number of complete duplicates 
is much less as compared to the number of zero eigenvalues, indicating that the contribution to occurrence of the zero eigenvalues comes mainly from the partial duplicates. 
With an increase in the average degree, the number of duplicate nodes as well as the number of zero eigenvalues decrease.
In order to further explore the impact of duplicates on zero degeneracy, we consider model networks other than the ER random networks generated using different 
algorithms, some of which might support node duplication. In case of a SF network of the same size and the average degree, a much higher 
number of duplicates and zero eigenvalues are exhibited as compared to the ER random network (Fig.~\ref{0_vs_model} (a)), but both the counts decrease with 
an increase in the average degree. 
The configuration model networks display less complete duplicates and zero eigenvalues values than that of the SF networks, as apart from the degree sequence there 
does not exist any other preference for association of nodes. With an increase in the average degree, number of complete duplicates and the zero eigenvalues decrease. On a further increase in the average degree, the number of complete duplicates and zero eigenvalues coincide to negligible values for all the three networks. 
At a fixed average degree, with increase in the size of the networks, both number of complete duplicates and zero eigenvalues increase in case of SF and configuration model networks, with zero degeneracy being higher as compared to complete duplication. The number of duplicates and zero eigenvalues however, remain negligible in case of ER random networks even with increase in size (Fig.~\ref{0_vs_model} (b)).

Keeping in view the high zero degeneracy prevalent in real world systems \cite{zero_real}, in the following we attempt to analyze how our investigation pertaining to model networks shed light on to the reasons behind high degeneracy at zero in real world systems. We analyze the protein-protein interaction (PPI) networks of six different species namely {\it H. Pylori}, {\it H. sapiens}, {\it D. melanogaster}, {\it S. cerevisiae}, {\it C. elegans} 
and {\it E. coli}. As depicted in Table~\ref{table2}, the number of zero eigenvalues are more than the number of complete duplicates indicating the existence of partial duplicates in the underlying networks. 
We generate ER random networks of the same size and average degree as of the six PPI networks. Table~\ref{table2} reveals that the generated ER random networks have absolutely no degeneracy at zero eigenvalues and no 
duplicates.
 As the PPI networks are SF in nature \cite{phya2014}, we generate SF networks of the same size and average degree as of the PPI networks using BA algorithm. We find that 
though corresponding SF networks lead to a high degeneracy at zero, as expected the number of zero eigenvalues and complete duplicates are much less than those of corresponding PPI networks. It may be due to the fact that the SF networks so generated display power law behaviour and need not have the same degree sequence as of the PPI networks. We further construct corresponding randomized models of the real systems, i.e. configuration models having the same degree sequence as of the six PPI networks. We observe that the configuration model networks have much higher zero degeneracy and complete duplication as compared to the SF networks generated using BA algorithm. This observation is quite intriguing as it has been demonstrated that the SF networks generated using the BA algorithm have higher duplication and degeneracy as compared to their corresponding configuration models (Fig.~\ref{lambdaER_SF_k} and \ref{0_vs_model}). But it would be noteworthy to mention that the configuration models generated in the case of model networks were the ones having exactly the same degree sequence as of the BA-algorithm generated SF networks. While in case of the PPI networks, the configuration models preserve the degree sequence of the PPI networks. This indicates that not only the power law behaviour of the networks accounts for duplication, but the very nature by which the real world PPI networks have evolved and acquires a degree sequence which favors duplication and leads to degeneracy at zero.

To conclude, all the real world networks show more zero degeneracy than the corresponding random models, indicating that equivalent number of nodes are completely duplicated (condition (a)) or have partial duplications (condition (b)), as depicted by Table~\ref{table2}. Gene duplication mechanism has been emphasized in evolutionary biology to be a driving force for creating new genes in a genome \cite{Teichmann,Hurles,prokaryotes,eukaryotes}. As duplicated genes are known to acquire mutations faster than other genes resulting in divergence of functions \cite{Duplication}, the PPI networks exhibiting prevalence of duplicates is quite interesting as well as intriguing.

Most of the real world networks are scale-free in nature, which renders few nodes connected with almost all other nodes in the network, leaving lot of nodes having as less as one connection with the hub only. This naturally leads to lot of complete duplicate nodes (condition (a)), which in turn leads to a high degeneracy at zero. Scale free networks generated through preferential attachment algorithm yield a good number of duplicated nodes, owing to very much nature of the algorithm, in turn leading to equivalent degeneracy at the zero eigenvalue. Since in the preferential attachment, a new entry in the network has more probability to connect with the existing hubs making it more probable that it connects with the same set of nodes which gained connections with the previous entry. Since such kind of biases does not exist in configuration model, as expected it exhibits less duplicates and hence less degeneracy at zero.
With an increase in the average degree, it becomes difficult for pair(s) of nodes to satisfy condition (a) or (b), as more connections will lead to more probability of destroying duplication, leading to a constant decrease in the zero eigenvalues.

We explore the origin of zero degeneracy in the network spectra. Our analysis sheds light on the mechanisms which collectively lead to zero degeneracy in the real networks. 
Further, we correlate the occurrence of zero degeneracy with the evolutionary origin of a network. Comparison of the number of duplicates and zero degeneracy in the PPI networks of six different species with their corresponding configuration models reveals that in addition to the power law behaviour of the real networks, 
other factors also contribute to node duplication leading to comparatively high zero degeneracy. Duplicated gene pairs have been emphasized to confer evolutionary stability to many biological systems \cite{Kitano_2004,Nowak}. The analysis carried out in this work combined with the occurrence of exceptionally high peak at zero degeneracy in real world networks, can be extended to understand other complex systems as well as to build up robust technological networks.

{\it Acknowledgements}:
SJ and AY thank Council of Scientific and Industrial
Research (CSIR), Govt. of India for the grant 25(0205)/12/EMR-II and for the fellowship, respectively. AY thanks Sanjiv Kumar Dwivedi and Camellia Sarkar of the Complex Systems Lab for fruitful discussions.

\end{document}